\newcommand{\coeff}{a}
\newcommand{\D}{A}
\newcommand{\obs}{B}
\newcommand{\B}{H}
\newcommand{\dW}{d\Gamma}
\newcommand{\Za}{Z_{\rm a}}
\newcommand{\Ea}{E_{\rm a}}
\newcommand{\Wa}{W_{\rm a}}
\newcommand{\Ed}{E_{\rm d}}
\newcommand{\xid}{\xi_{\rm d}}
\newcommand{\hd}{h_{\rm d}}

\newcommand{\kB}{k_{\rm B}}
\newcommand{\kT}{k_{\rm B}T}
\newcommand{\Pa}{\frac{1-S_{2}}{3}}
\newcommand{\Ra}{\frac{S_{4}}{35}-\frac{2S_{2}}{21}+\frac{1}{15}}
\newcommand{\Rb}{\frac{1}{7}(S_{2} - S_{4})}

\newcommand{\fij}{f_{ij}}
\newcommand{\G}{G}
\newcommand{\wij}{\omega_{ij}}
\newcommand{\wkl}{\omega_{kl}}
\newcommand{\Gij}{{\bf G}_{ij}}
\newcommand{\Gjk}{{\bf G}_{jk}}
\newcommand{\I}{{\bf 1}}
\newcommand{\Ei}{{\bf \Gamma}_{i}}
\newcommand{\Ej}{{\bf \Gamma}_{j}}
\newcommand{\Ek}{{\bf \Gamma}_{k}}
\newcommand{\Di}{{\bf \Lambda}_{i}}
\newcommand{\Ga}{{\bf \Gamma}}
\newcommand{\Gb}{{\bf \Lambda}}

\newcommand{\Qls}{{\cal C}}
\newcommand{\Rls}{\bar{\cal R}}
\newcommand{\Rns}{{\cal R}}
\newcommand{\Sls}{{\cal S}}
\newcommand{\Tls}{{\cal T}}
\newcommand{\Uls}{{\cal U}}
\newcommand{\Vls}{{\cal V}}

\newcommand{\h}{\vec{h}}
\newcommand{\e}{\vec{s}}
\newcommand{\n}{\vec{n}}
\newcommand{\ei}{\vec{s}_{i}}
\newcommand{\ej}{\vec{s}_j}

\newcommand{\nii}{\vec{n}_{i}}
\newcommand{\nj}{\vec{n}_j}

\newcommand{\rij}{\vec{r}_{ij}}
\newcommand{\vij}{\hat{r}_{ij}}

\newcommand{\la}{\left\langle}
\newcommand{\ra}{\right\rangle}
\newcommand{\raa}{\right\rangle_{\rm a}}
\newcommand{\qikjl}{q_{ik:jl}}
\newcommand{\qiljk}{q_{il:jk}}

\documentstyle[eqsecnum,aps,prb,floats,epsfig]{revtex}
\begin{document}

\title{
Thermodynamic perturbation theory for dipolar superparamagnets
}
\author{
P. E. J{\"o}nsson and J. L. Garc\'{\i}a-Palacios\cite{leave}
}
\address{
Department of Materials Science, Uppsala University
\\
Box 534, SE-751 21 Uppsala, Sweden
}
\date{\today}
\maketitle

\begin{abstract}
Thermodynamic perturbation theory is employed to derive analytical
expressions for the equilibrium linear susceptibility and specific
heat of lattices of anisotropic classical spins weakly coupled by the
dipole-dipole interaction.
The calculation is carried out to the second order in the coupling
constant over the temperature, while the single-spin anisotropy is
treated exactly.
The temperature range of applicability of the results is, for weak
anisotropy ($\D/\kT\ll1$), similar to that of ordinary high-temperature
expansions, but for moderately and strongly anisotropic spins
($\D/\kT\gtrsim1$) it can extend down to the temperatures where the
superparamagnetic blocking takes place ($\D/\kT\sim25$), provided only
the interaction strength is weak enough.
Besides, taking exactly the anisotropy into account, the results
describe as particular cases the effects of the interactions on
isotropic ($\D=0$) as well as strongly anisotropic ($|\D|\to\infty$)
systems (discrete orientation model and plane rotators).
\end{abstract}


\section{Introduction}

In paramagnetic systems the relative weakness of the 
dipole-dipole interaction between
magnetic ions results in characteristic temperatures lying in the
range of $0.01$--$0.1$\,K.
Besides, the dipolar coupling usually coexists with the exchange
interaction.
However, for superparamagnets (nanoscale solids or clusters whose net
spin rotates thermally activated in the magnetic anisotropy
potential), the exchange or other competing interactions can usually
be discarded and the dipole-dipole interaction can be studied in pure
form.
In addition, the size of their typical magnetic moments
($S\sim10^{2}$--$10^{5}$) shifts the relevant temperatures up to the
range of a few Kelvin, facilitating greatly the experimental study of
this interaction.\cite{Luo91jonetal95}

The calculation of the relevant statistical-mechanical quantities
constitutes a formidable problem in most many-body systems.
Apart from various specific solution ansatzs, a number of systematic
expansions (in the density, coupling parameter, etc.) have been
developed for weak interactions.\cite{balescu}
In spin and dipole systems, the moment method of Van Vleck
\cite{vanvle37,vanvle40} (a high-temperature expansion of the
partition function) permits to study the equilibrium properties in
the absence of cooperative phenomena.
This technique is one of the few analytical tools available to handle
systems coupled by the dipole-dipole interaction, due to the
long-range and reduced symmetry of this interaction.

An important property of superparamagnets is their magnetic
anisotropy, which results in a number of spin orientations of minimum
energy separated by potential barriers.
For uniaxial spins \cite{uniaxial} the characteristic time for the
thermo-activated rotation of the spin over the anisotropy barrier $\D$
can approximately be written as
\begin{equation}
\label{arrhenius:tau}
\tau
\simeq
\tau_{0}\exp(\D/\kT)
,
\end{equation}
where $\tau_{0}$ is weakly temperature dependent and takes values
$\tau_{0}\sim10^{-10}$--$10^{-12}$\,s for magnetic nanoparticles.
Then, for a given measurement time, $t_{\rm m}$, the system exhibits
its thermal-equilibrium response when the {\em condition of
superparamagnetism}, $t_{\rm m}\gg\tau$, is obeyed, which corresponds
to the temperature range (in units of $\D/k_{\rm B}$):
\begin{equation}
\label{superparamagnetic-range}
\ln(t_{\rm m}/\tau_{0})>\D/\kT\geq 0
.
\end{equation}
In ``static" measurements ($t_{\rm m}\sim1$--$100$\,s), due to the
small value of $\tau_{0}$, this equilibrium range extends down to very
low temperatures ($25>\D/\kT$), showing that the naive ascription of
superparamagnetism to the range in which ``the thermal energy is
comparable or larger than the anisotropy energy" ($1\gtrsim \D/\kT$)
is unduly restrictive.
Indeed, as $T$ decreases the system displays different behaviors
ranging from almost isotropic ($\D/\kT\ll1$), then moderately
anisotropic ($\D/\kT\sim1$), and eventually strongly anisotropic
($\D/\kT\gg1$) {\em without leaving the equilibrium regime}.
Therefore, descriptions based on the assumption of isotropic behavior
or the opposite discrete-orientation or plane-rotator approximations
(for easy-axis and easy-plane anisotropy), necessarily have a reduced
range of validity in superparamagnets.

Due to the mentioned characteristics of the dipole-dipole coupling and
the difficulties introduced by the anisotropy, most rigorous
calculations in interacting superparamagnets have been done by
numerical simulation techniques.\cite{zalcie93andetal97bergor2001}
In this article we apply thermodynamic perturbation theory 
\cite{TPT} to calculate analytically the linear
susceptibility and the specific heat of lattices of uniaxial classical
spins coupled by the dipole-dipole interaction, accounting exactly
(non-perturbatively) for their anisotropy energy.
Along with the study of the dependence on the shape of the system of
certain quantities (due to the long-range of the dipole-dipole
interaction \cite{gri68bangriwid98}), our treatment permits to
investigate the effects of the strength and sign of the anisotropy, as
well as of the orientational distribution of anisotropy axes.

We find that for systems with axes {\em oriented at random} the
corrections to the specific heat and the linear susceptibility become
independent of the anisotropy (at least to second order in the
interaction coupling) in certain spatial arrangements of the spins
(e.g., cubic or completely disordered).
The latter is a generalization to interacting systems of the well known
absence of anisotropy effects on the equilibrium linear susceptibility of dipoles with random anisotropy (discussed in Ref.\ \onlinecite{vanvle40}).
However, apart from the important exception of random axes, the
anisotropy is an essential element in the determination of the
corrections due to the interactions.
This is illustrated with the response of systems with {\em parallel\/}
anisotropy axes, where an ordinary high-temperature expansion, either
disregarding the anisotropy or in the discrete orientation limit,
poorly describes the susceptibility curves (computed for comparison by
Monte Carlo simulation), while the thermodynamic perturbation theory
describes the results with reasonable accuracy.

\section{Thermodynamic perturbation theory for interacting dipoles}

In this section we introduce the spin system studied and discuss the
application of perturbation theory to calculate approximately
thermodynamic quantities.

\subsection{Hamiltonian of a system of interacting anisotropic spins}

Let us consider a system of $N$ magneto-anisotropic spins coupled by
the dipole-dipole interaction.
The magnetic anisotropy energy is assumed to be uniaxial
\begin{equation}
\Ea
=
-\D\sum_{i}(\ei \cdot \nii)^2
\;,
\end{equation}
where $\D$ is the anisotropy parameter (for magnetic nanoparticles
$\D=KV$, where $K$ and $V$ are the anisotropy constant and volume), and
$\ei$ and
$\nii$ are, respectively, unit vectors along the magnetic moment and
anisotropy axis of the $i$th spin.
The dipole-dipole interaction
energy can be written as
\begin{equation}
\label{dipolar}
\Ed
=
-\frac{\mu_{0}m^{2}}{4\pi a^{3}}
\sum_{i>j} \wij
\;,
\qquad
\wij
=
\ei \cdot \Gij \cdot \ej
\;,
\end{equation}
where $m$ is the magnitude of the magnetic moment, $a$ is an
appropriate characteristic length (see below), and
\begin{eqnarray}
\label{Gij}
\Gij
&=&
\frac{1}{r_{ij}^{3}}
\left(3\,\vij \vij - \I\right),
\\
\rij
&=&
\vec{r}_{i}-\vec{r}_j,
\quad
\vij
=
\rij/r_{ij}
\;.
\end{eqnarray}
Here $\I$ is the unit tensor and $\rij$ the vector joining the sites
$i$ and $j$ (measured in units of $a$).
%
The action of a tensor dyadic ${\bf T}=\vec{u}\,\vec{v}$ on
a vector $\vec{w}$ is the usual one
$(\vec{u}\,\vec{v})\vec{w}
\equiv
\vec{u}(\vec{v}\cdot\vec{w})$,
and hence the tensor $\Gij$, when multiplied with $\ej$, gives (except
for a constant) the field at the position of the $i$th dipole created
by $\ej$.

For notational simplicity we are assuming that the parameters
characterizing the different dipoles are identical, but it is immediate
to generalize the expressions for different anisotropy constants,
magnetic moments, volumes, etc.
Similarly, although the assumption of uniaxial anisotropy is not
necessary until the end of the calculation we made it here for
definiteness.
Concerning the characteristic length $a$, it is defined in such a way
that $a^3$ is the mean volume around each spin.
Thus, $a$ is the lattice constant in a simple cubic arrangement, and
for nanoparticles the volume concentration of particles is $c=V/a^3$.

Finally, introducing $\beta=1/\kT$ and the following dimensionless
quantities (anisotropy and coupling constant relative to the thermal
energy)
\begin{equation}
\label{dimless}
\sigma
=
\frac{\D}{\kT}
\;,
\qquad
\xid
=
\frac{\mu_{0}m^2}{4\pi a^3}
\frac{1}{\kT}
\;,
\end{equation}
we can write the total energy $E=\Ea+\Ed$ as
\begin{equation}
-\beta E
=
\sigma \sum_{i}(\ei \cdot \nii)^2
+
\xid
\sum_{i>j}\wij
\;.
\end{equation}
Note that the interaction strength can also be measured by the
temperature independent coupling parameter
\begin{equation}
\label{hdip}
\hd
=
\xid/2\sigma
\;,
\end{equation}
which is the magnitude of the field (measured in units of the maximum
anisotropy field $\mu_{0}\B_{K}=2\D/m$) produced at a given position
by a dipole at a distance $a$.\cite{hd&c}

\subsection{Equilibrium linear susceptibility and specific heat}

The thermal-equilibrium average of any quantity
$\obs(\e_{1},\ldots,\e_{N})$ is given by
\begin{equation}
\label{avgH0}
\la\obs\ra
=
\frac{1}{Z}
\int\!
\dW
\obs
\,
\exp (-\beta E)
\;,
\end{equation}
where $Z = \int\!\dW\exp (-\beta E)$
is the partition function.
In classical spins the different states correspond to different spin
orientations so that $\dW=\prod_{i}d\Omega_{i}$, with
$d\Omega_{i}
=
d^{2}\ei/2\pi$.

The linear susceptibility is defined as the derivative of the
magnetization $\frac{1}{N}\langle\sum_{i}\ei\rangle$ with respect to
the {\em external\/} magnetic field (which is the experimentally
manipulable quantity in contrast to the internal macroscopic field).
However, from basic statistical mechanics we know that the response to
a probing field $\Delta\vec{\B}$ can be obtained in terms of suitable
averages of the net spin taken in the absence of $\Delta\vec{\B}$.
If in addition there is no external {\em bias\/} field applied, the
susceptibility is simply given by
\begin{equation}
\label{susceptibility}
\chi
=
\frac{\mu_{0}m^{2}}{\kT}
\frac{1}{N}
\big\langle s_{z}^2 \big\rangle
,
\qquad
s_{z}=\sum_{i}( \ei\cdot\h )
,
\end{equation}
where $\h$ is a unit vector along $\Delta\vec{\B}$ and $s_{z}$ is the
field projection of the net moment.
The specific heat at constant volume $c_{v}=\partial\langle
E\rangle/\partial T$ can be obtained directly from $Z$ as
\begin{equation}
\frac{c_{v}}{\kB}
=
\beta^2\frac{\partial^2}{\partial \beta^2}(\ln Z)
=
\sigma^2\frac{\partial^2}{\partial \sigma^2}(\ln Z)
\;,
\label{cV}
\end{equation}
where to take the $\sigma$-derivative, the coupling parameter
$\xid$ is expressed as $\xid=2\sigma\hd$ [Eq.\ (\ref{hdip})].
As in the calculation of $\chi$, we only consider the
zero-field specific heat.\cite{garpal2000acp}

\subsection{Thermodynamic perturbation theory}

We shall now use thermodynamic perturbation theory, \cite{TPT} 
to expand the Boltzmann 
distribution $W=Z^{-1}\exp (-\beta E)$ in powers of $\xid$.
This will lead to an expression of the form
\begin{equation}
\label{W:approx}
W
=
\Wa
\left(
1
+
\xid
F_{1}
+\case{1}{2}
\xid^{2}
F_{2}
+
\cdots
\right)
\;,
\end{equation}
where $F_{1}$ is linear in $\Ed$ (and hence quadratic in the spins),
$F_{2}$ is up to quadratic in $\Ed$ (quartic in the $\ei$), and
\begin{equation}
\label{W:non-int}
\Wa
=
\Za^{-1}
\exp(-\beta\Ea)
\;,
\end{equation}
is the Boltzmann distribution of the noninteracting ensemble.
Therefore, the calculation of an observable $\la\obs\ra$ is reduced to
the calculation of averages weighted by $\Wa$ (denoted $\langle \cdot
\rangle_{\rm a}$) of typically low grade powers of the spin variables:
$\langle\obs\rangle_{\rm a}$, $\langle\obs F_{1}\rangle_{\rm a}$,
$\langle\obs F_{2}\rangle_{\rm a}$, etc.
An ordinary high-temperature expansion corresponds to expand
Eq.\ (\ref{W:non-int}) further in powers of $\beta=1/\kT$.
In the present calculation, however, the averages are kept weighted
over $\Wa$, so they will be exact in the magnetic anisotropy and only
perturbational in the dipolar interaction.\cite{previousTPT}

A convenient way of performing the expansion in
powers of $\xid$ is to introduce the Mayer functions $f_{ij}$
defined by $1+\fij=\exp(\xid\wij)$, which permits to write the
exponential in the Boltzmann factor as
\begin{equation}
\exp(-\beta E)
=
\exp(-\beta \Ea)\prod_{i>j}(1+\fij).
\label{expBoltz}
\end{equation}
Expanding the product to second order in the $\fij$ gives
\begin{equation}
\label{prod:fij}
\prod_{i>j}(1+\fij)
=
1
+
\xid
\G_{1}
+
\case{1}{2}
\xid^2
\G_{2}
+
O(\xid^3)
,
\label{expfij}
\end{equation}
where\cite{ros-lax52}
\begin{eqnarray}
\label{g1}
\G_{1}
&=&
\sum_{i>j}\wij,
\\
\label{g2}
\G_{2}
&=&
\sum_{i>j} \wij^2 + \sum_{i>j} \sum_{k>l} \wij \wkl \qikjl \qiljk,
\end{eqnarray}
and the symbol $\qikjl$ annihilates terms containing duplicate pairs:
$\qikjl =
\case{1}{2}
(2-\delta_{ik}-\delta_{jl})
(1+\delta_{ik})(1+\delta_{jl})$.

To obtain the average of any quantity $\obs$ we introduce the
expansion (\ref{prod:fij}) in both the numerator and denominator of
$\la\obs\ra
=
\int\!\dW\obs\,\exp (-\beta E)
/\int\!\dW\exp (-\beta E)$,
and work out the expansion of the quotient, getting\cite{noteF}
\begin{eqnarray*}
\la\obs\ra
&\simeq&
\la\obs\ra_{\rm a}
+
\xid
\big[
\la\obs\,\G_{1} \ra_{\rm a}
-
\la\obs\ra_{\rm a} \la \G_{1} \ra_{\rm a}
\big]
\\
& &
{}+\case{1}{2}
\xid^2
\Big\{
\la\obs\,\G_{2} \ra_{\rm a}
- \la\obs\ra_{\rm a} \la \G_{2} \ra_{\rm a}
\\
& &
\qquad\quad
{}- 2\la \G_{1} \ra_{\rm a}
\big[
\la\obs\,\G_{1} \ra_{\rm a} 
-
\la\obs\ra_{\rm a} \la \G_{1}\ra_{\rm a}
\big]
\Big\}
\;.
\end{eqnarray*}
However, since in our case $\Wa(-\ei)=\Wa(\ei)$ (because
the single-spin anisotropy has inversion symmetry and there is no bias
field) and in $\G_{1}=\sum_{i>j}\wij$ a dipole does not interact with
itself, the result $\la \G_{1}\ra_{\rm a}=0$ holds.
Under these conditions we finally find the simpler form
\begin{eqnarray}
\la\obs\ra
&\simeq&
\la\obs\ra_{\rm a}
+
\xid
\la\obs\,\G_{1} \ra_{\rm a}
\nonumber\\
& &
{}+\case{1}{2}
\xid^2
\big[
\la\obs\,\G_{2} \ra_{\rm a}
-
\la\obs\ra_{\rm a} \la \G_{2} \ra_{\rm a}
\big].
\label{avg:g1=0}
\end{eqnarray}

To complete the calculation we need to obtain averages of low grade
powers of $\e$ weighted by the noninteracting distribution (moments),
which is the only place where one needs to specify the form of $\Ea$.
We can write the susceptibility and specific heat to second order in
$\xid$ using up to fourth order moments, which are calculated in
Appendix \ref{App:alg} [Eqs.\ (\ref{alg2}) and (\ref{alg4})] for the
uniaxial distribution $\Wa\propto\exp[\sigma(\e\cdot\n)^2]$.
For instance, the first two moments read
$\la(\vec{c}_{1}\cdot\e)\ra_{\rm a}=0$ and
\[
\la ( \vec{c}_{1} \cdot \e)(\vec{c}_{2} \cdot \e) \ra_{\rm a}
=
\Pa \, \vec{c}_{1} \cdot \vec{c}_{2}
+ S_{2} (\vec{c}_{1} \cdot \vec{n})(\vec{c}_{2} \cdot \vec{n})
\;,
\]
where the $\vec{c}_{n}$ are arbitrary constant vectors and $S_{l}$ is
the average of the $l$th Legendre polynomial $P_{l}(z)$ over $\Wa$:
\begin{equation}
\label{Sl}
S_l(\sigma)
=
\la P_{l}(\e\cdot\n)\ra_{\rm a}
\;.
\end{equation}
In particular, $S_{0}=1$ and
$S_{2}(\sigma)
=
\frac{1}{2}\la 3(\vec{s} \cdot \vec{n})^{2}-1 \ra_{\rm a}$
can be expressed in terms of familiar special functions (error
functions or the Dawson integral), while other $S_{l}$ appearing in
higher order moments, can be obtained by means of the recurrence
relation they satisfy [Eq.\ (\ref{Xlm:EOM:uniaxial:m=0:stat})].

\section{Analytical expressions for the susceptibility and
specific heat}

We shall now present the general form of the perturbative expressions
for the susceptibility and specific heat and discuss their properties
in some important cases.

\subsection{Linear susceptibility}

Using the results of the previous section, we have averaged the square
of the field projection of the net spin
[$\obs=(\sum_{i}\ei\cdot\h)^{2}]$, which yields an expansion for the
equilibrium linear susceptibility [Eq.\ (\ref{susceptibility})] of the
form
\begin{equation}
\label{eqsusc}
\chi
=
\frac{\mu_{0}m^{2}}{\kT}
\left(
\coeff_{0}
+
\xid
\coeff_{1}
+
\case{1}{2}
\xid^2
\coeff_{2}
\right)
\;.
\end{equation}
The lengthy general expressions for the coefficients $\coeff_{n}$,
which include sums over the lattice of the dipolar tensor $\Gij$ and
the $S_l$, are written in full in Appendix \ref{App:genform}.

The coefficients $\coeff_{n}$ simplify notably for some orientational
distributions of the anisotropy axes.
For systems with parallel axes (e.g., single crystals of magnetic
molecular clusters, or a ferrofluid frozen in a strong field), the
coefficients for the longitudinal response read
\begin{eqnarray}
\label{a0para}
\coeff_{0,\parallel}
&=&
\frac{1+2S_{2}}{3}
\\
\label{a1para}
\coeff_{1,\parallel}
&=&
\frac{1+4S_{2}+4S_{2}^2}{9}
\,
\Qls
\\
\label{a2para}
\case{1}{2}
\coeff_{2,\parallel}
&=&
-\frac{1+4S_{2}+4S_{2}^2}{27}
\nonumber\\
& &
{}\times
\left[
(1-S_{2}) \, \big( \Rls - \Sls\big)
+
3S_{2} \, \left( \Tls - \Uls \right)
\right]
\nonumber\\
& &
{}+ \frac{7 + 10S_{2}-35S_{2}^2+18S_{4}}{315}
\nonumber\\
& &
{}\times
\left[
(1-S_{2}) \, \Vls
+
3S_{2} \, \big(\Tls - \case{1}{3}\Rls\big)
\right]
\;,
\end{eqnarray}
where $\Qls$, $\Rns$ ($\Rls$), $\Sls$, $\Tls$, $\Uls$, and $\Vls$ are
certain lattice sums whose properties are discussed below.

To obtain the susceptibility when the anisotropy axes are distributed
at random, we average the general expressions for the $\coeff_{n}$
over $\n$, with help from Eqs.\ (\ref{alg2:iso}) and (\ref{alg4:iso})
(with $\e\to\n$), getting
\begin{eqnarray}
\label{a0ran}
\coeff_{0,{\rm ran}}
&=&
\frac{1}{3}
\\
\label{a1ran}
\coeff_{1,{\rm ran}}
&=&
\frac{1}{9}
\,
\Qls
\\
\label{a2ran}
\case{1}{2}
\coeff_{2,{\rm ran}}
&=&
-\frac{1}{27}
\,
\big( \Rls - \Sls \big)
+
\frac{1}{45}\,
(1-S_{2}^2)\, \Vls
\;.
\end{eqnarray}
Note that in the limit of isotropic spins (where $S_{l}\to0$) the
results for coherent axes and for random anisotropy duly coincide.

\subsection{Specific heat}

To obtain the specific heat we can expand directly the partition
function in powers of $\xid$ by introducing the expanded Boltzmann
factor [Eqs.~(\ref{expBoltz}) and (\ref{expfij})] in the definition of
$Z$
\begin{eqnarray}
Z
&=&
\int \dW \exp(-\beta E_a)
(1 +\xid \G_1 + \case{1}{2} \xid^2 \G_2)
\nonumber
\\
&=& 
\Za (1 + \case{1}{2} \xid^2 \la \G_2 \raa)
\label{Zexp}
\;,
\end{eqnarray}
where the linear term vanishes because $\la \G_1 \raa=0$.
Then $c_{v}$ is obtained [Eq.~(\ref{cV})] by differentiating the
logarithm of the above expansion, which poses no problem of the type
of expanding a quotient, since to second order in $\xid$ we can use
$\ln(1+x\xid^{2})\simeq x\xid^{2}$.
The result has the form
\begin{equation}
\label{cVexp}
\frac{c_{v}}{N \kB}=\sigma^2 b_0 + \case{1}{2} \xid^2 b_2
\;,
\end{equation}
where the zeroth order coefficient
\begin{equation}
\label{b0gen}
b_0
=
\frac{4}{315}(18S_4 -35S_2^2+10S_2+7)
\;,
\end{equation}
gives the specific heat in the absence of interactions.

The general formula for $b_{2}$ is given in Appendix \ref{App:cV}
[Eq.\ (\ref{b2gen})].
Again, it simplifies for coherent axes and for random anisotropy.
In the first case ($\nii=\n$, $\forall i$) we obtain
\begin{eqnarray}
b_{2,\parallel}
&=&
\case{1}{3}
\Big\{
1-S_2^2
-
4\sigma S_2 S_2' 
-
\sigma^2 [S_2 S_2'' + (S_2')^2]
\Big\} 
\Rns
\nonumber \\
& &
+\case{1}{3}
\Big(
2S_2(1-S_2)
+
4\sigma S_2'(1-2S_2) 
\nonumber\\
& &
\qquad
{}+
\sigma^2 \{S_2''-2[S_2 S_2'' + (S_2')^2]\}
\Big)
\Vls
\nonumber \\
&&+
\Big\{
S_2^2
+
4 \sigma S_2 S_2'
+
\sigma^2 [S_2S_2'' + (S_2')^2]
\Big\}
\Tls
\;,
\end{eqnarray} 
where $f'=df/d\sigma$.
For randomly distributed axes, on averaging the general expression for
$b_2$ over $\n$ by means of Eq.~(\ref{alg2:n}), one gets
\begin{equation}
b_{2,{\rm ran}}=\case{1}{3}\cal{R}.
\end{equation}
This is the same correction term as that obtained for {\em
isotropic\/} spins by Waller \cite{wal36} and Van Vleck.
\cite{vanvle37}

\subsection{The lattice sums}

An essential element of the expressions derived for $\chi$ and $c_{v}$
are the following ``lattice sums''
\begin{eqnarray}
\label{Qsum}
\Qls
&=&
\frac{1}{N}\sum_{i} \sum_{j \ne i}
\h\cdot\Gij\cdot\h
\\
\label{Rsum}
\Rns
&=&
\frac{2}{N}\sum_{i} \sum_{j \ne i}
r_{ij}^{-6}
\\
\label{Rpsum}
\Rls
&=&
\frac{1}{N}\sum_{i} \sum_{j \ne i}
\h\cdot\Gij\cdot \Gij\cdot\h
\\
\label{Ssum}
\Sls
&=&
\frac{1}{N}\sum_{i} \sum_{j \ne i} \sum_{k \ne j}
\h\cdot\Gij\cdot\Gjk\cdot\h
\\
\label{Tsum}
\Tls
&=&
\frac{1}{N}\sum_{i} \sum_{j \ne i}
(\h\cdot\Gij\cdot\h)^2
\\
\label{Usum}
\Uls
&=&
\frac{1}{N}\sum_{i} \sum_{j \ne i} \sum_{k \ne j}
(\h\cdot\Gij\cdot\h)(\h\cdot\Gjk\cdot\h)
\\
\label{Vsum}
\Vls
&=&
\frac{1}{N}\sum_{i} \sum_{j \ne i}
r_{ij}^{-3} \h\cdot\Gij\cdot\h
\end{eqnarray}
(replace $\h$ by $\n$ in the formulas for $c_{v}$).
Considering the structure of these sums along with the form of the
dipolar tensor $\Gij$ [Eq.\ (\ref{dipolar})], some physical
interpretation can be provided for the different terms in the
perturbative series.\cite{ros-lax52}
The first order term $\coeff_{1}$ incorporates through $\Qls$ the {\em
direct\/} action ($j\to i$) on each spin of the remainder spins, when
aligned along the probing field.
No term of this type appears in the specific heat since the absence of
any external field yields $b_{1}\equiv0$.
The second order terms $a_{2}$ and $b_{2}$ involve lattice sums
including products of $\Gij$ with $\Gjk$ so they take into account the
action on a given spin of the others but through intermediate spins
({\em indirect\/} action $k\to j\to i$).
In particular, if $k=i$ we have the {\em reaction\/} on the $i$th
spin of its direct action on the remainder spins.

In the next section we shall compute $\chi$ and $c_{v}$ for
``sufficiently isotropic" lattices, in the sense of fulfilling
$\sum(r_x)^n=\sum(r_y)^n=\sum(r_z)^n$, e.g., cubic and completely
disordered lattices (incidentally, the type of arrangements for which
in the classical Lorentz cavity-field calculation the contribution of
the dipoles inside the ``small sphere'' vanishes).
In these lattices we have two important results: (i) $\Rls$ coincides
with the more familiar lattice sum $\Rns$ (which justifies the
notation) and (ii) $\Vls=0$.

\subsection{Ordinary high-temperature expansions and other
approximations}

The effects of the anisotropy are included {\em exactly\/} in our
formulas through the anisotropy-weighted averages of the Legendre
polynomials $S_{2}(\sigma)$ and $S_{4}(\sigma)$, which modulate the
different contributions and introduce an extra dependence on the
temperature via $\sigma=\D/\kT$.
This reflects the fact that the dipolar field at a given site is
different if the source spins are almost freely rotating (high $T$,
$S_{l}\to0$) or, for example, lay almost parallel to their respective
anisotropy axes (low $T$, $S_{l}\to1$).

The approximate behavior of $S_{2}$ and $S_{4}$ for weak
($|\sigma|\ll1$) and strong ($|\sigma|\gg1$) anisotropy is
\cite{S2S4approx}
\begin{eqnarray}
\label{S2exp}
S_{2}(\sigma)
&=&
\left\{ \begin{array}{ll}
\frac{2}{15}\sigma+\frac{4}{315}\sigma^2 + \cdots
&
|\sigma| \ll 1
\\
1 - \frac{3}{2\sigma}-\frac{3}{4\sigma^2} + \cdots
&
\sigma \gg 1
\\
-\frac{1}{2}(1+\frac{3}{2\sigma}) +\cdots
&
\sigma \ll -1
\end{array}
\right.,
\\
\label{S4exp}
S_{4}(\sigma)
&=&
\left\{
\begin{array}{ll}
\frac{4}{315}\sigma^2 + \cdots
&
|\sigma| \ll 1
\\
1 - \frac{5}{\sigma}+\frac{25}{4\sigma^2} + \cdots
&
\sigma \gg 1
\\
\frac{3}{8}(1+\frac{5}{\sigma}+\frac{35}{4\sigma^2}) + \cdots
&
\sigma \ll -1
\end{array}
\right..
\end{eqnarray}
The results of an ordinary high temperature expansion, in which all
terms in $\exp(-\beta E)$ are expanded, correspond to replace $S_{2}$
and $S_{4}$ in the coefficients $\coeff_{n}$ and $b_{n}$ by their weak
anisotropy approximations.
Besides, taking the $\sigma\to0$ limit (where $S_{l}=0$) we get the
known results for isotropic spins (in a slightly more general form,
since the terms including $\Vls$ are usually omitted due to the
lattices assumed\cite{ros-lax52,zhobag92}).
In addition, substituting the above strong anisotropy formulas in
$\chi$ and $c_{v}$ we get these quantities in the discrete-orientation
and plane-rotator cases (with the corresponding corrections in powers
of $1/\sigma$).

\section{Behavior of the susceptibility and the specific heat}

In this section we study the features of the susceptibility and
specific heat emerging from the analytical expressions derived.
We shall discuss the shape dependence of these quantities, investigate
the dependence on the anisotropy (on both its strength and the axes
distribution), and finally estimate the limits of validity of the
expansions.
For concreteness, we shall consider the behavior of spins with
easy-axis anisotropy ($\D>0$) in simple cubic lattices.

\subsection{Shape dependence}

Here we restrict our attention to systems with ellipsoidal shape,
which is the geometry usually consider in studies of the dipole-dipole
interaction because: (i) in the continuous limit the demagnetizing
field is spatially homogeneous and parallel to the external field
when this is along one of the three principal axes
\cite{stratton41} and (ii) it covers as limit cases important
geometries such as those of disks or long cylinders.

The long range of the dipole-dipole interaction leads to a shape
dependence of the physical quantities in an external
field\cite{gri68bangriwid98} and hence of the linear susceptibility
(which is a field derivative).
In the expressions obtained, this shape dependence is borne by
the slowly convergent lattice sums $\Qls$, $\Sls$, and $\Uls$.
For the class of sufficiently isotropic lattices mentioned, these sums
vanish in macroscopically large {\em spherical\/} systems, being
nonzero otherwise.
The sums $\Rns$ ($\Rls$), $\Tls$, and $\Vls$, on the other hand,
contain $r_{ij}^{-6}$ (instead of $r_{ij}^{-3}$ or
$r_{ij}^{-3}r_{jk}^{-3}$), which makes them rapidly convergent and
shape-independent.\cite{values}

Figure \ref{Fig:shapedep} shows the thermodynamic perturbation theory
susceptibility (thin lines) for $\hd=\xid/2\sigma=0.02$ and $\chi$ in
the noninteracting limit (thick lines) for a system with parallel axes
with the form of a prolate ellipsoid, a sphere, and an oblate ellipsoid
(the symbols correspond to Monte Carlo simulations to be discussed
below).
For the non spherical systems $\Qls\neq0$,\cite{continuous} and the
corrections to the susceptibility are largely dominated by the first
order term, which corresponds to the direct action discussed above.
The shape dependence is easily understood by recalling the behavior of
two dipoles: if their axes are aligned, they minimize their
dipole-dipole energy lying parallel along the line joining them, while
if this line is perpendicular to the axes, they minimize the
interaction energy pointing along opposite directions.
Therefore, in elongated systems the aligning effect dominates and
$\chi$ is larger than the noninteracting susceptibility [Fig.\
\ref{Fig:shapedep}(a)], while in oblate systems the opposite occurs
with the associated decrease of $\chi$ [Fig.\ \ref{Fig:shapedep}(c)].
In the sphere the direct term exactly cancels ($\Qls=0$), and negative
second order corrections (which incorporate indirect and reaction
terms) determine the susceptibility [Fig.\
\ref{Fig:shapedep}(b)].

Concerning the specific heat, due to the presence of the rapidly
convergent lattice sums $\Rns$, $\Tls$, and $\Vls$, this quantity does
not depend on the shape of the system.
Physically, this is a consequence of the absence of linear term in
Eq.\ (\ref{Zexp}), which follows from $\la\G_1 \raa=0$, which in turn
requires the absence of a bias field.

\subsection{Anisotropy dependence}

To illustrate the importance of taking the anisotropy into account, we
are going to compare the thermodynamic perturbation theory results
with: (i) those obtained by an ordinary high temperature expansion for
zero anisotropy (the $S_{l}\to0$ limit of our expressions) and (ii)
the results in the discrete orientation limit ($S_{l}\to1$).

In the cases of cubic or completely disordered lattices
we already know that the lattice sum $\Vls$ vanishes.
Then, the first corrections to the susceptibility due to the
interactions become {\em exactly\/} independent of the anisotropy {\em
if\/} the axes are distributed at random [see Eqs.\
(\ref{a0ran})--(\ref{a2ran}) where the only anisotropy dependent term
is multiplied by $\Vls$].
Therefore, the susceptibility coincides with that obtained by an
ordinary high-temperature expansion for isotropic dipoles (see Ref.\
\onlinecite{ros-lax52}; $\Vls=0$ was implicitly used in that
work) but also with the perturbative $\chi$ for Ising-like spins with
randomly distributed ``Ising'' axes [Fig.\ \ref{Fig:rp}(a)].
This generalizes to interacting spins the well known result, discussed
in Ref.\ \onlinecite{vanvle40} (see also Ref.\
\onlinecite{garpaljonsve2000}), i.e., the absence of anisotropy effects
on the equilibrium linear susceptibility of systems with random
anisotropy.

Nevertheless, when $\Vls\neq0$ (as occurs in tetragonal lattices) the
susceptibility will show some anisotropy dependence for random axes
(although weak, since $\Vls\neq0$ is accompanied by $\Qls\neq0$ and
$\chi$ is then dominated by the anisotropy independent first order
term).
In any case, on inspecting Eqs.\ (\ref{a0para})--(\ref{a2para}) one
would expect important differences for parallel axes between the
thermodynamic perturbation theory results and those were the
anisotropy is not included ($S_{l}\to0$).
This is what actually occurs as Fig.\ \ref{Fig:rp}(b) illustrates: the
susceptibility is not only larger for parallel axes than for isotropic
dipoles, but also the temperature dependence is stronger, owing to the
extra, anisotropy-induced, temperature dependence of the coefficients
$\coeff_{n}$ via $S_{l}(\sigma)$.
This is clearly seen when comparing the correction terms $\Delta\chi$
themselves [inset of Fig.\ \ref{Fig:rp}(b)].

Note that the susceptibilities in the isotropic and Ising cases
constitute lower and upper bounds to the actual $\chi$.
The upper bound is slowly approached at low enough temperatures
($\sigma\gg1$), completing in this way the crossover from isotropic
behavior at high $T$ to the discrete-orientation behavior at low $T$.
Note finally that the lowest temperatures displayed ($\sigma\sim20$) are
still inside the range where an ordinary magnetization experiment
yields the equilibrium response ($\sigma\sim25$).

\begin{figure}[htb]
\centerline{\epsfig{figure=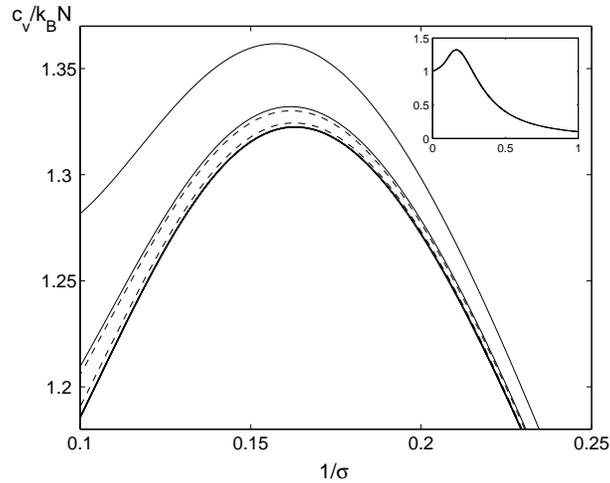,width=8cm}}
\caption{
Specific heat per spin vs.\ temperature for non-interacting spins
(thick line), and weakly interacting spins with randomly distributed
anisotropy axes (dashed lines) and parallel axes (thin lines) arranged
on a simple cubic lattice.
In each case, $\hd=\xid/2\sigma=0.003$ and $0.006$ from bottom to top.
The inset shows the specific heat for non-interacting spins over a
wider temperature interval.
}
\label{Fig:cv}
\end{figure}

Concerning the specific heat, the part corresponding to the
noninteracting system, $b_0$, does not depend on the anisotropy axes
orientations.
The reason is that if the spins are independent, they cannot probe the
relative orientations of their axes and there is no preferential
direction to compare their orientations with (as that of the probing
field in the susceptibility).
Thus, the noninteracting specific heat (inset in Fig.~\ref{Fig:cv})
only reflects the individual behavior of the spins in their
single-spin anisotropy potentials, and its peak (at $\sigma\sim5$)
reflects the ``transition'' from the isotropic behavior at high $T$
(with $c_{v}\propto1/T^{2}$) to the discrete orientation behavior at
low $T$.
This can be considered as a sort of Schottky peak, due in this case
to the ``depopulation" of the high-energy ``barrier levels''.\cite{cv0}

The corrections due to the coupling depend naturally on the
orientations of the axes.
For a given axes distribution the specific heat increases with the
interaction strength (Fig.\ \ref{Fig:cv}), since an amount of heat
injected in the system can partially be stored in the form of the
potential energy of interaction.
For the same reason, since the average of the interaction energy is
larger (in magnitude) in systems with aligned axes, one would expect a
larger $c_{v}$ in this case.
Figure~\ref{Fig:cv} shows that the effect of the interaction is indeed
stronger in a system with aligned axes than in a system with random
anisotropy.

\subsection{Validity limits of the perturbational results}

Since the analytic expressions derived for $\chi$ and $c_{v}$ are
expansions valid in principle for $\xid\ll1$, which corresponds to
$T\propto1/\sigma\gg2\hd$ [Eq.\ (\ref{hdip})], the results will
deviate appreciably from the exact quantities at sufficiently low
$T$.
In order to estimate the limits of validity, it would be desirable to
compare the formulas with the results of some method treating the
interactions without approximations.
Besides, as the fundamentals of the expansion are the same for the
different quantities, it would suffice to estimate the range of
validity for one of them.

We have compared the analytical $\chi$ with the susceptibility
computed by a Monte Carlo method (described in Appendix \ref{App:MC}),
which except for statistical and finite sampling errors is exact to
compute equilibrium properties.
Returning to Fig.\ \ref{Fig:shapedep} we observe that $\chi$ obtained
by thermodynamic perturbation theory (thin lines) describes accurately
the simulated susceptibility (symbols) at high and intermediate $T$,
while the results start to deviate slightly at the lowest temperatures
displayed ($\sigma\sim4$).
Therefore, since $\hd=0.02$ was used in this graph, an estimate of the
lower temperatures attainable is $\xid\sim1/6$, which is milder than
the a priori restriction $\xid\ll1$.

\section{Summary and Conclusions}

We have obtained approximate analytical expressions for the
equilibrium linear susceptibility and specific heat of classical spins
interacting via dipole-dipole interactions by means of thermodynamic
perturbation theory.
The formulas account for the interactions to second order in the
coupling constant over temperature but are exact in the anisotropy.
As the results are valid for any strength and sign of the anisotropy,
provided only the interaction strength is weak enough
($\xid\lesssim1/6$), they include as particular cases the linear
response and specific heat of isotropic as well as strongly
anisotropic spins (discrete-orientation model and plane rotators).

The expressions derived also account for the different orientational
distributions of anisotropy axes.
For randomly distributed axes and sufficiently isotropic lattices
(e.g., cubic or completely disordered), the linear susceptibility
becomes independent of the anisotropy, at least to second order in the
coupling constant.
This extends to interacting systems the well known absence of
anisotropy effects on the equilibrium linear response of systems with
random anisotropy.
The same holds for the corrections to the specific heat due to the
interactions (indeed, without restrictions on the lattice type).
For a general axes distribution, however, the anisotropy effects do
not disappear.
The importance of including them has been illustrated in the case of
coherent axes by showing the failure of ordinary high-temperature
expansions (for either isotropic or strongly anisotropic spins) to
describe the exact susceptibility in cases where the thermodynamic
perturbation theory yields accurate results.

\acknowledgments

We acknowledge F.\ J.\ L\'azaro for useful discussions and The Swedish
Natural Science Research Council (NFR) for financial support.

\appendix

\section{Averages weighted with the uniaxial anisotropy
Boltzmann factor}
\label{App:alg}

The averages we need to calculate are all of products of the form $I_m
=
\left\langle
\prod_{n=1}^{m}(\vec{c}_n \cdot \e)
\right\rangle_{\rm a}$,
where the $\vec{c}_{n}$ are arbitrary constant vectors.
Introducing the polar and azimuthal angles of the spin,
$(\vartheta,\varphi)$, we can write $I_{m}$ as
\[
I_m
=
\frac{\int_0^{2\pi} d\varphi \int_0^{\pi} d\vartheta \sin\vartheta
\prod_{n=1}^{m}(\vec{c}_n \cdot \e) \exp[\sigma(\e\cdot\n)^2]}
{\int_0^{2\pi} d\varphi \int_0^{\pi} d\vartheta \sin\vartheta
\exp[\sigma(\e\cdot\n)^2]}
\;.
\]
For odd $m$, $I_m$ is an integral of an odd function over a
symmetric interval and hence $I_m=0$.
To calculate the susceptibility and specific heat to second order in
$\xid$, we require $I_{2}$ and $I_4$, which will be
calculated using symmetry arguments similar to those employed to
derive the $\sigma=0$ unweighted averages (see, for instance,
Ref.~\onlinecite{mathews-walker}).

Note that $I_{2}$ is a scalar bilinear in $\vec{c}_{1}$ and
$\vec{c}_{2}$.
The most general scalar with this property that can be constructed
with the vectors of the problem ($\vec{c}_{1}$, $\vec{c}_{2}$, and
$\n$) has the form
\[
I_{2}
=
A \, \vec{c}_{1} \cdot \vec{c}_{2}
+
B \,(\vec{c}_{1} \cdot \n)(\vec{c}_{2} \cdot \n)
\;.
\]
To find the coefficients $A$ and $B$ one chooses particular values for
the $\vec{c}_n$: (i) If $\vec{c}_{1} \parallel
\vec{c}_{2} \perp \n$ then $I_{2} = A$.
Thus, setting $\n=\hat{z}$ and $\vec{c}_{1}=\vec{c}_{2}=\hat{x}$, one
has $\e\cdot\n=\cos\vartheta\equiv z$ and $( \vec{c}_{1} \cdot
\e)(\vec{c}_{2} \cdot \e)=(1-z^{2})\cos^2\varphi$, so the integral reads
\begin{eqnarray*}
A
&=&
\frac{
\int_0^{2\pi} d\varphi \cos^2\varphi
\int_{-1}^{1} dz
(1-z^{2})
\exp(\sigma z^{2})
}{
\int_0^{2\pi} d\varphi
\int_{-1}^{1} dz
\exp(\sigma z^{2})
}
\\
&=&
\case{1}{2}
\big[
1-\big\langle z^{2}\big\rangle_{\rm a}
\big]
=
\Pa
\;,
\end{eqnarray*}
where $S_{2}(\sigma)=\la P_{2}(z) \ra_{\rm a}$ is the average of the
second Legendre polynomial
$P_{2}(z)=\frac{1}{2}\left(3\,z^{2}-1\right)$ over the noninteracting
distribution.
(ii) If $\vec{c}_{1}\parallel\vec{c}_{2}\parallel\n$, then
$I_{2}=A+B$.
Putting $\n=\vec{c}_{1} = \vec{c}_{2} = \hat{z}$ the integral is given
by
\[
A+B
=
\frac{
\int_{-1}^{1} dz\,
z^{2}
\exp(\sigma z^{2})
}
{
\int_{-1}^{1} dz
\exp(\sigma z^{2})
}
=
\big\langle
z^{2}
\big\rangle_{\rm a}
=
\frac{1+2S_{2}}{3}
\;.
\]
Therefore, since
$I_{2}=\la ( \vec{c}_{1} \cdot \e)(\vec{c}_{2} \cdot \e)\ra_{\rm a}$,
we get for the second order moment
\begin{equation}
\la ( \vec{c}_{1} \cdot \e)(\vec{c}_{2} \cdot \e) \ra_{\rm a}
=
\Pa \, \vec{c}_{1} \cdot \vec{c}_{2}
+ S_{2} (\vec{c}_{1} \cdot \vec{n})(\vec{c}_{2} \cdot \vec{n})
\;.
\label{alg2}
\end{equation}
We can similarly calculate $I_4$ by constructing the
most general scalar fulfilling certain properties, getting
\begin{eqnarray}
& &
\langle
(\vec{c}_{1} \cdot \e)
(\vec{c}_{2} \cdot \e)
(\vec{c}_{3} \cdot \e)
(\vec{c}_{4} \cdot \e)
\rangle_{\rm a}
\nonumber\\
&=&
\Delta_{4} \,
[ (\vec{c}_{1} \cdot \vec{c}_{2})(\vec{c}_{3} \cdot \vec{c}_{4})+
(\vec{c}_{1} \cdot \vec{c}_{3})(\vec{c}_{2} \cdot \vec{c}_{4})+
(\vec{c}_{1} \cdot \vec{c}_{4})(\vec{c}_{2} \cdot \vec{c}_{3}) ]
\nonumber\\
& &
{}+\Delta_{2} \,
[(\vec{c}_{1} \cdot \vec{c}_{2})(\vec{c}_{3} \cdot \vec{n})
(\vec{c}_{4} \cdot \vec{n})+
(\vec{c}_{1} \cdot \vec{c}_{3})(\vec{c}_{2} \cdot \vec{n})
(\vec{c}_{4} \cdot \vec{n})
\nonumber\\
& &
\qquad\quad
{}+(\vec{c}_{1} \cdot \vec{c}_{4})(\vec{c}_{2} \cdot \vec{n})
(\vec{c}_{3} \cdot \vec{n})+
(\vec{c}_{2} \cdot \vec{c}_{3})(\vec{c}_{1} \cdot \vec{n})
(\vec{c}_{4} \cdot \vec{n})
\nonumber\\
& &
\qquad\quad
{}+(\vec{c}_{2} \cdot \vec{c}_{4})(\vec{c}_{1} \cdot \vec{n})
(\vec{c}_{3} \cdot \vec{n})+
(\vec{c}_{3} \cdot \vec{c}_{4})(\vec{c}_{1} \cdot \vec{n})
(\vec{c}_{2} \cdot \vec{n})]
\nonumber\\
& &
{}+ S_{4} (\vec{c}_{1} \cdot \vec{n})(\vec{c}_{2} \cdot \vec{n})
(\vec{c}_{3} \cdot \vec{n})(\vec{c}_{4} \cdot \vec{n}),
\label{alg4}
\end{eqnarray}
where $\Delta_{2}$ and $\Delta_{4}$ are combinations of the first
$S_l(\sigma)$
\begin{equation}
\label{deltas}
\Delta_{2}
=
\Rb
,
\qquad
\Delta_{4}
=
\Ra
\;.
\end{equation}
Therefore, Eq.\ (\ref{alg4}) involves $S_{2}$ as well as
$S_{4}(\sigma)=\la P_{4}(z)\ra_{\rm a}$, the average of the fourth
Legendre polynomial
$P_{4}(z)=\frac{1}{8}\left(35\,z^{4}-30\,z^{2}+3\right)$ with respect
to $\Wa$.

Finally, introducing the following tensor and scalar shorthands
\begin{eqnarray}
\label{tensorGa}
\Ga
&=&
\Pa \I + S_{2} \, \n\,\n
\;,
\\
\label{tensorGb}
\Gb
&=&
\sqrt{\Delta_{4}}\, \I + \frac{\Delta_{2}}{\sqrt{\Delta_{4}}} \,\n\,\n
\;,
\quad
\Omega
=
S_{4} - 3\frac{\Delta_{2}^2}{\Delta_{4}}
\;,
\end{eqnarray}
where $\I$ is the identity tensor, the results for the moments can
compactly be written as
\begin{eqnarray}
\la ( \vec{c}_{1} \cdot \e)(\vec{c}_{2} \cdot \e) \ra_{\rm a}
&=&
(\vec{c}_{1} \cdot \Ga \cdot \vec{c}_{2})
\\
\langle
(\vec{c}_{1} \cdot \e)
(\vec{c}_{2} \cdot \e)
(\vec{c}_{3} \cdot \e)
(\vec{c}_{4} \cdot \e)
\rangle_{\rm a}
&=&
(\vec{c}_{1} \cdot \Gb \cdot \vec{c}_{2})
(\vec{c}_{3} \cdot \Gb \cdot \vec{c}_{4})
\nonumber\\
& &
{}+(\vec{c}_{1} \cdot \Gb \cdot \vec{c}_{3})(\vec{c}_{2} \cdot \Gb \cdot
\vec{c}_{4})
\nonumber\\
& &
{}+(\vec{c}_{1} \cdot \Gb \cdot \vec{c}_{4})(\vec{c}_{2} \cdot \Gb \cdot
\vec{c}_{3})
\nonumber\\
& &
{}+\Omega (\vec{c}_{1} \cdot \n)(\vec{c}_{2} \cdot \n)
(\vec{c}_{3} \cdot \n)(\vec{c}_{4} \cdot \n)
\;,
\end{eqnarray}
which facilitates the manipulation of the
observables.

The quantities $S_{l}$ can be computed using the following homogeneous
three-term recurrence relation\cite{kalcof97}
\begin{equation}
\label{Xlm:EOM:uniaxial:m=0:stat}
\bigg[
1
-
\frac{2\sigma}{(2l-1)(2l+3)}
\bigg]
S_{l}
-
\frac{2\sigma}{2l+1}
\bigg[
\frac{l-1}{2l-1}
S_{l-2}
-
\frac{l+2}{2l+3}
S_{l+2}
\bigg]
=
0
\;,
\end{equation}
knowing the first two terms: $S_0=1$ and $S_2$, which is given by
\begin{eqnarray}
\label{S2}
S_{2}
&=&
\frac{3}{2}
\left(
\frac{e^{\sigma}}{\sigma \Za}
-
\frac{1}{2\sigma}
\right)
-\frac{1}{2}
\;.
\end{eqnarray}
The one-spin partition function $\Za=\int_{-1}^{1}d z\,\exp(\sigma
z^{2})$ can be written in terms of {\em error\/} functions of real and
``imaginary" argument as
\begin{equation}
\label{Z}
\Za
=
\left\{
\begin{array}{ll}
\sqrt{\pi/\sigma}\,{\rm erfi}(\sqrt{\sigma})
,
&
\sigma>0
\\
\sqrt{\pi/|\sigma|}\,{\rm erf}(\sqrt{|\sigma|})
,
&
\sigma<0
\end{array}
\right.
\end{equation}
The less familiar ${\rm erfi}(x)$ is related with the Dawson integral
$D(x)$, so in the easy-axis case one can write
$\Za=(2e^{\sigma}/\sqrt{\sigma})D(\sqrt{\sigma})$ and compute $D(x)$
with the subroutine {\tt DAWSON} of Ref.\ \onlinecite{recipes}.

Note finally that in the isotropic limit ($S_l\to0$), Eqs.~(\ref{alg2})
and (\ref{alg4}) reduce to the known moments for the isotropic
distribution \cite{ros-lax52,mathews-walker}
\begin{eqnarray}
\label{alg2:iso}
\langle ( \vec{c}_{1} \cdot \e)(\vec{c}_{2} \cdot \e)
\rangle_{{\rm iso}}
&=&
\case{1}{3} \, \vec{c}_{1} \cdot \vec{c}_{2}
\;,
\\
\label{alg4:iso}
\langle
(\vec{c}_{1} \cdot \e)
(\vec{c}_{2} \cdot \e)
(\vec{c}_{3} \cdot \e)
(\vec{c}_{4} \cdot \e)
\rangle_{{\rm iso}}
&=&
\case{1}{15}
[
(\vec{c}_{1} \cdot \vec{c}_{2})(\vec{c}_{3} \cdot \vec{c}_{4})
\nonumber\\
& &
\qquad
{}+
(\vec{c}_{1} \cdot \vec{c}_{3})(\vec{c}_{2} \cdot \vec{c}_{4})
\nonumber\\
& &
\qquad
{}+
(\vec{c}_{1} \cdot \vec{c}_{4})(\vec{c}_{2} \cdot \vec{c}_{3})]
\;.
\end{eqnarray}
These expressions are formally identical to those for the
average of a quantity involving the anisotropy axes $\nii$, when
these are distributed at random
$\frac{1}{N}
\sum_{i}
f(\nii)
\rightarrow
\int
\frac{d^{2}\n}{4\pi}
f(\n)
\equiv
\overline{f}$.
For instance, for arbitrary $\n$-independent vectors $\vec{v}_{1}$ and
$\vec{v}_{2}$, we have
\begin{equation}
\label{alg2:n}
\frac{1}{N}
\sum_{i}
(\vec{v}_{1}\cdot\nii)(\vec{v}_{2}\cdot\nii)
\longrightarrow
\overline{(\vec{v}_{1}\cdot\n)(\vec{v}_{2}\cdot\n)}
=
\case{1}{3} \, \vec{v}_{1}\cdot\vec{v}_{2}
\;.
\end{equation}

\section{General formulae for the coefficients of the susceptibility}
\label{App:genform}

The general expression for the equilibrium linear susceptibility is
given by Eq.~(\ref{eqsusc}) with the following expressions for the
coefficients
\begin{eqnarray*}
\coeff_{0}
&=&
\frac{1}{N}\sum_{i} \h\cdot\Ei\cdot\h
\\
\coeff_{1}
&=&
\frac{1}{N}\sum_{i} \sum_{j \ne i}
\h\cdot\left(\Ei \cdot \Gij \cdot \Ej\right)\cdot\h
\\
\coeff_{2}
&=&
- \frac{2}{N} \sum_{i} \sum_{j \ne i}
\h\cdot\left(\Ei \cdot \Gij \cdot \Ej \cdot \Gij \cdot \Ei\right)\cdot\h
\\
& &
{}+
\frac{2}{N} \sum_{i} \sum_{j \ne i}\sum_{k \ne j}
\h\cdot\left( \Ei \cdot \Gij \cdot \Ej \cdot \Gjk \cdot \Ek \right)
\cdot\h
\\
& &
{}+ \frac{1}{N}\sum_{i} \sum_{j \ne i}
\bigg\{
\frac{1-S_{2}}{r_{ij}^{6}}
\Big[ (\h\cdot\Di\cdot\h)(\vij \cdot \Di \cdot \vij)
\\
& &
\qquad\qquad\qquad
\qquad\qquad
{}+2(\h\cdot\Di \cdot \vij)^2
\\
& &
\qquad\qquad\qquad
\qquad\qquad
{}+
\Omega (\h\cdot\nii)^2(\nii \cdot \vij)^2\Big]
\\
& &
\qquad\qquad\qquad
{}+S_{2}
\Big[
(\h\cdot\Di\cdot\h)
(\nj \cdot \Gij \cdot \Di \cdot \Gij \cdot \nj)
\\
& &
\qquad\qquad\qquad
\qquad\quad
{}+
2 (\h\cdot\Di \cdot \Gij \cdot \nj)^2
\\
& &
\qquad\qquad\qquad
\qquad\quad
{}+
\Omega (\h\cdot\nii)^2(\nii \cdot \Gij \cdot \nj)^2
\Big]
\bigg\}
\\
& &
{} - \frac{1}{N}\sum_{i} \sum_{j \ne i} (\h\cdot\Ei\cdot\h)
\bigg[
\frac{1-S_{2}}{r_{ij}^{6}}
(\vij \cdot \Ei \cdot \vij)
\\
& &
\qquad\qquad\qquad
\qquad\qquad\quad
{}+S_{2} (\nj \cdot \Gij \cdot \Ei \cdot \Gij \cdot \nj)
\bigg]
\end{eqnarray*}
where $\Gij$, $\rij$ and $\vij$ are defined in Eq.\ (\ref{Gij}), and
$\Ga$, $\Gb$, and $\Omega$ in Eqs.\ (\ref{tensorGa})--(\ref{tensorGb})
and also involve the $S_l(\sigma)$.

When calculating these coefficients, the same type of averages appear as
in the isotropic case (see Refs.~\onlinecite{vanvle37,ros-lax52} for
details of the calculation) and with the same multiplicities.
The only difference is the weight function and hence the formulas
required to calculate those averages [Eqs.~(\ref{alg2}) and
(\ref{alg4}) instead of Eqs.~(\ref{alg2:iso}) and (\ref{alg4:iso})].

\section{General formula for the coefficient $b_2$ of the specific heat}
\label{App:cV}

In the general expression (\ref{cVexp}) for the specific heat the
coefficient $b_{0}$ is given by Eq.\ (\ref{b0gen}), while $b_{2}$ reads
\begin{eqnarray}
Nb_2 
&=&
\case{1}{3} \big\{ 2(1-S_2) -4\sigma S_2' -\sigma^2S_2'' \big\}
 \sum_{i}\sum_{j \ne i} r_{ij}^{-6} 
\nonumber \\
& &
+\case{1}{2}
\big\{
2S_2(1-S_2) + 4\sigma S_2'(1-2S_2)
\nonumber\\
& &
\qquad
{}+ \sigma^2 [S_2''(1-2S_2) - 2 (S_2')^2]
\big\} 
\nonumber\\
& &
\quad
\times
\sum_{i} \sum_{j \ne i} r_{ij}^{-6}
\big[(\vij \cdot \nii)^2+(\vij \cdot \nj)^2\big]
\nonumber \\
& &
+
\big\{ S_2^2 + 4\sigma S_2 S_2' + \sigma^2 [S_2S_2'' + (S_2')^2] \big\}
\sum_{i}\sum_{j \ne i} (\nii \cdot \Gij \cdot \nj)^2 \; ,
\label{b2gen}
\end{eqnarray}
where $f'=df/d\sigma$.
An arbitrary $S_{l}'$ can be expressed in terms of averaged Legendre
polynomial by means of the following differential-recurrence relation
\begin{eqnarray}
\label{dSldsig}
\frac{d S_{l}}{d\sigma}
&=&
\frac
{(l-1)l}
{(2l-1)(2l+1)}
S_{l-2}
+
\frac
{2l(l+1)}
{3(2l-1)(2l+3)}
S_{l}
\nonumber\\
& &
{}+
\frac
{(l+1)(l+2)}
{(2l+1)(2l+3)}
S_{l+2}
-
\frac{2}{3}
S_{2}
S_{l}
\;.
\end{eqnarray}
This useful formula can readily be demonstrated by taking the
derivative of the definition of
$S_{l}
\equiv
\langle P_{l}\rangle_{\rm a}
=
\int_{-1}^{1}\!dz\,P_{l}\,e^{\sigma z^{2}}
/
\int_{-1}^{1}\!dz\,e^{\sigma z^{2}}$:
\begin{eqnarray}
\label{dSldsig:1}
\frac{dS_{l}}{d\sigma}
&=&
\frac
{
\int_{-1}^{1}\!dz\,
z^{2}
P_{l}\,
e^{\sigma z^{2}}
}
{
\int_{-1}^{1}\!dz\,
e^{\sigma z^{2}}
}
-
\frac
{
\int_{-1}^{1}\!dz\,
P_{l}\,
e^{\sigma z^{2}}
\int_{-1}^{1}\!dz\,
z^{2}
e^{\sigma z^{2}}
}
{
\big(
\int_{-1}^{1}\!dz\,
e^{\sigma z^{2}}
\big)^{2}
}
\nonumber\\
&=&
\langle
z^{2}
P_{l}
\rangle_{\rm a}
-
\case{1}{3}
S_{l}
\left(1+2 S_{2}\right)
\;,
\end{eqnarray}
where $\langle z^{2} \rangle_{\rm a}=(1+2 S_{2})/3$ has been used.
The product of $P_{l}$ with $z$ can be expanded in Legendre
polynomials by using the corresponding relation for associated
Legendre functions $P_{l}^{m}$ (Ref.\ \onlinecite{arfken},
Ch.\ 12)
\begin{equation}
\label{Legendre:relation}
z P_{l}
=
\frac{1}{2l+1}
\left[
l P_{l-1}
+
(l+1)P_{l+1}
\right]
\;.
\end{equation}
Multiplying $zP_{l}$ by $z$, using again Eq.\
(\ref{Legendre:relation}) to expand $zP_{l\pm1}$ on the right-hand
side, and gathering terms yields
\begin{eqnarray*}
z^{2} P_{l}
&=&
\frac
{(l-1)l}
{(2l-1)(2l+1)}
P_{l-2}
+
\frac
{2l(l+1)-1}
{(2l-1)(2l+3)}
P_{l}
\\
& &
\quad
{}+
\frac
{(l+1)(l+2)}
{(2l+1)(2l+3)}
P_{l+2}
\;.
\end{eqnarray*}
Averaging this result and substituting in Eq.\ (\ref{dSldsig:1}), we
finally get the desired Eq.\ (\ref {dSldsig}).

\section{Monte Carlo simulations}
\label{App:MC}

In order to obtain nonperturbative results to test the analytical
expressions we have performed careful Monte Carlo simulations, by a
method similar to that employed in Ref.~\onlinecite{nowak2000}.
The trial Monte Carlo step (MCS) is a random rotation of the spin
within a cone, achieved by generating a random vector with uniform
probability on the surface of a sphere of radius $\varrho$, adding the
random vector to the initial spin, and finally normalizing the
resulting vector.
For a better acceptance ratio of the generated configurations, we
scale that radius with $T$ as $\varrho=0.7/\sqrt{\sigma}$.

In terms of the maximum anisotropy field $\mu_{0}\B_{K}=2\D/m$,
where $\D$ is the anisotropy parameter and $m$ is the magnetic moment,
a dimensionless probing field can be defined as $\Delta\B/\B_{K}$ or,
in temperature units, as $\Delta\xi=2\sigma\Delta\B/\B_{K}$.
The latter is the argument in the Langevin function for the (isotropic)
magnetization, and it controls if the response is linear in $\Delta\B$.
Therefore, to treat all the temperatures on the same footing
$\Delta\xi\propto\Delta\B/T$ is kept constant in the simulation
($\Delta\xi=0.15$) which requires to decrease the probing field with
$T$.
The value $\Delta\xi=0.15$ used should ensure linear response since it
is well below the values $\Delta\xi=0.3$--$0.5$, where nonlinear terms
can start to contribute appreciably to the equilibrium response
(see, for instance, Appendix C of Ref.\
\onlinecite{garpallaz98}).

\begin{figure}[htb]
\centerline{\epsfig{figure=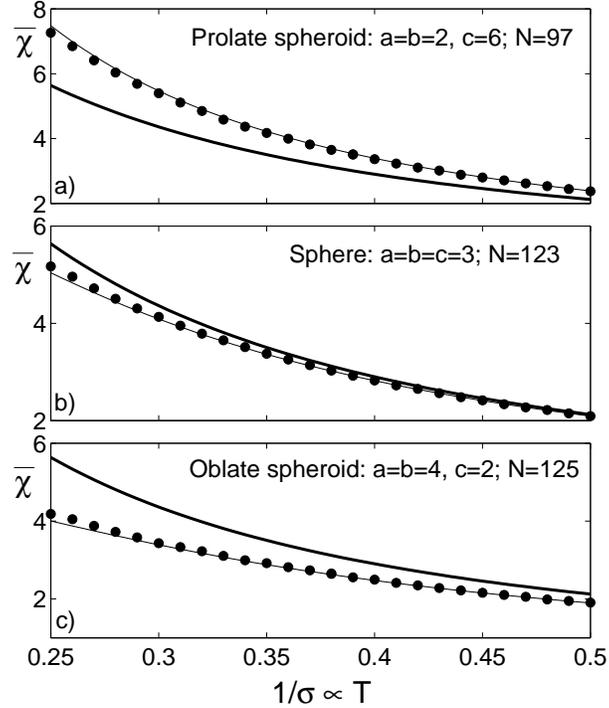,width=8cm}}
\caption{
Equilibrium linear susceptibility vs.\ temperature for three different
ellipsoidal systems with equation $x^2/a^2+y^2/b^2+z^2/c^2\leq1$
resulting in a system of $N$ dipoles.
The susceptibility is given in reduced units
$\bar{\chi}=\chi(\B_{K}/m)$, the spatial arrangement of the spins is
simple cubic and the probing field is applied along the anisotropy
axes, which are {\em parallel\/} to the $z$-axis.
The thick lines are the equilibrium susceptibility of the
corresponding noninteracting systems (they are equal in all three
cases); thin lines are the susceptibilities including the corrections
due to the dipolar interactions obtained by thermodynamic perturbation
theory [Eq.\ (\ref{eqsusc})]; and the symbols represent the
susceptibility obtained with a Monte Carlo method.
The dipolar interaction strength is $\hd=\xid/2\sigma=0.02$.
(For the prolate and oblate ellipsoids the second order correction is
very small in the temperature interval displayed and omitting it
the curves visually coincide.)}
\label{Fig:shapedep}
\end{figure}

\begin{figure}[htb]
\centerline{\epsfig{figure=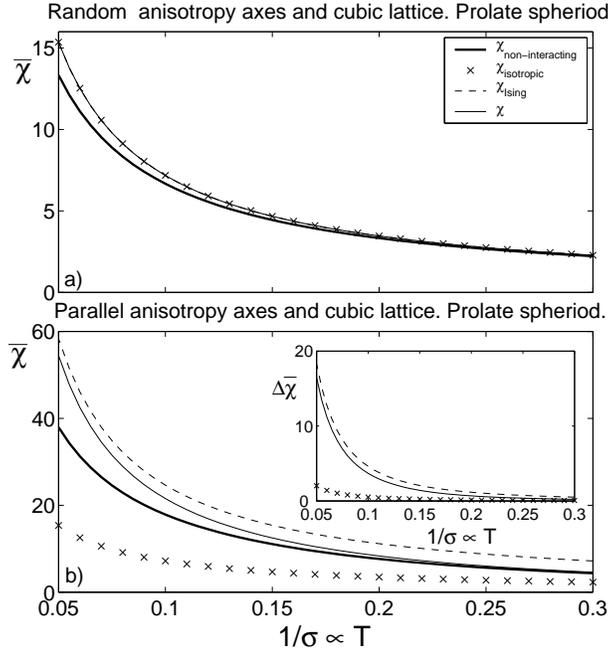,width=8cm}}
\caption{
Equilibrium linear susceptibility vs.\ temperature for the
same prolate ellipsoid as in Fig.\ \ref{Fig:shapedep}(a) and the
dipolar interaction strength $\hd=\xid/2\sigma=0.004$.
The spins are arranged on a simple cubic lattice with: a) randomly
distributed anisotropy axes and b) parallel anisotropy axes.
Thick lines are the susceptibilities for independent spins and thin
lines are the susceptibilities obtained by thermodynamic perturbation
theory.
For comparison we have displayed $\chi$ obtained by a classical
high temperature expansion for isotropic spins (crosses) and for Ising
spins (dashed line). 
Inset: Comparison of the corrections to the susceptibility due to
interactions ($\Delta\chi=\chi-\chi_{\rm non-int}$).
(Note that the temperatures displayed are above the lower validity
limit $\xid\sim1/6$ estimated in the text:
$1/\sigma=2\hd/\xid\sim0.048$.)}
\label{Fig:rp}
\end{figure}

The number of dipoles in each simulation is indicated in the panels of
Fig.\ \ref{Fig:shapedep} and the simulations are done with open
boundary conditions.
%
%
The absolute value of the coefficient in front of Eq.\ (\ref{dipolar})
can be written as $2\D\,\hd$, so that $\hd$ [Eq.\ (\ref{hdip})] is
used as the input in the simulation.
In order to control that the response is the equilibrium one, we apply
a sinusoidal probing field of low frequency ($f=8\times 10^{-6}$
MCS$^{-1}$) and check that further reducing $f$ does not change the
results.
Besides, the 10 first periods are excluded from the average and the
susceptibility sampled over the 20 following periods.


\begin{thebibliography}{10}
\bibitem{hd&c}
For magnetic nanoparticles, we can write the coupling parameter $\hd$
in terms of the particle volume concentration $c$ as 
$\hd=(M_{\rm s}/4\pi\B_{K})c$, where $M_{\rm s}=m/V$ is the
spontaneous magnetization of the particles and $\B_{K}$ the maximum
anisotropy field.
\bibitem{noteF}
The functions $F_{1}$ and $F_{2}$ in the generic expansion
(\ref{W:approx}) are related with $\G_{1}$ and $\G_{2}$ [Eqs.\
(\ref{g1}) and (\ref{g2})] by
\begin{eqnarray*}
F_{1}
&=&
\G_{1}
-
\la \G_{1} \ra_{\rm a}
\;,
\\
F_{2}
&=&
\G_{2}
-
\la \G_{2} \ra_{\rm a}
-
2\la \G_{1} \ra_{\rm a}
\left[
\G_{1}
-
\la \G_{1} \ra_{\rm a}
\right]
\;.
\end{eqnarray*}
\bibitem{continuous}
For example, in the continuous limit
$\Qls=4\pi(\case{1}{3}-\lambda_{z})$, where $\lambda_{z}$ is the
demagnetization factor along the $z$-axis, which takes the value
$\lambda_{z}=1/3$ in a sphere.
\bibitem{cv0}
The specific heat does not attain the value $c_{v}=0$ at $T=0$
because the spins are classical.
\end{thebibliography}
\bibliographystyle{prsty}

\end{document}